# Economics Bulletin

# Stock market integration in the Latin American markets: further evidence from nonlinear modeling

Fredj JAWADI
*Amiens School of Management and EconomiX-University Paris Ouest Nanterre La Défense*

Nicolas MILLION  
*Banque de France*

Mohamed El hédi Arouri  
*Université dOrléans - LEO and EDHEC*

## Abstract

This article studies the financial integration between the six main Latin American markets and the US market in a nonlinear framework. Using the threshold cointegration techniques of Hansen and Seo (2002), we show significant threshold stock market linkages between Mexico, Chile and the US. Thus, the dynamics of these markets depends simultaneously on local and global risk factors. More importantly, our results show an on-off threshold financial integration process that is activated only when the stock price adjustment exceeds some level.

# 1. Introduction

Studies on the international integration of Latin American stock markets have recently gained momentum in the finance literature (Richards 1995, Bekaert and Harvey 1995, Heimonen 2002, and Carrieri *et al.* 2007) to cite just a few). This increase in interest and motivation is explained by a variety of reasons. Firstly, since the 1980s, these emerging markets have been widely seen as one of the most exciting and promising areas for investment, especially because they are expected to generate high returns and to offer good portfolio diversification opportunities. Secondly, financial liberalization has been largely implemented in several Latin American countries via ongoing structural adjustment programs. As a prerequisite to the financial liberalization processes, stabilization policies have been designed to ensure macro-economic stability, low inflation and reduced budget deficits. Finally, the recurrent financial crises of the last decades have stimulated the appetite of researchers to study Latin American markets. This is largely due to the fact that financial institutions and policymakers hope to measure the intensity of the interdependence between national stock markets and thus develop reliable contagion risk management tools.

Previous studies on the integration of Latin American markets with major markets (mainly the US market) have highlighted significant linkages between these stock markets, and the authors have explained these findings in various ways. Among the explanations, the first arguments are linked to financial market legislation, regimentation and dynamism (substantial deregulation and liberalization of capital markets across most countries, the internationalization of multinational companies and the activities of global investors). The second arguments concern fundamental economic feedback with respect to stock markets (business cycle synchronization, economic integration, and the evolution of macroeconomic variables and their effects on market co-movements). Other explanations are based on behavioral finance, implying possible linkages between the stock markets that arise notably from the interaction between international investors (contagion and the international transmission of crisis).

Nevertheless, although most previous studies have agreed on the presence of significant linkages between Latin American and US stock markets (Bekaert *et al.* 2005, and references therein), these findings should be treated with great caution. In fact, the previous studies used linear econometric techniques: linear regressions, causality tests, linear cointegration approaches, and multivariate GARCH models, which, unfortunately, limit the integration dynamics to linear and continuous models with constant speed over time. However, several recent studies have suggested that the integration dynamic is time-varying and that the nonlinear framework would be more appropriate and robust as a tool for reproducing the time-varying linkages between stock markets. In fact, the recent increase in the number of international investors, market liberalization and the financial crisis may have induced some persistence, asymmetry, irregularity and nonlinearity into the stock market integration process, thereby ruling out the linear framework (Bekaert and Harvey 1995, Masih and Masih 2001, Heimonen 2002, Barari 2003, and Carrieri *et al.* 2007).

This article contributes to the existing literature by using new nonlinear econometric techniques to investigate emerging stock market integration in order to reproduce the possible persistence, asymmetry and discontinuities that characterize the financial integration process. We studied the integration between the six major Latin American stock markets and the US stock market by applying the threshold cointegration techniques developed by Hansen and Seo (2002). This approach enabled us not only to check stock market integration in the presence of market frictions, but also to specify a time-varying integration process that is active per regime only when stock price deviations exceed a certain threshold.

The remainder of the article is organized as follows. Section II briefly presents the econometric methodology. The empirical results are discussed in section III and the last section presents the conclusions.

## 2. Threshold Financial Integration Modeling

In relation to the traditional linear framework, the main contribution of the threshold cointegration literature is to specify an on-off adjustment process that may be asymmetric, discontinuous and active by regime. This is due to a threshold effect given by investors and transaction cost heterogeneity for instance, which yields a nonlinear error-correction mechanism that is active as soon as the adjustment deviations exceed this threshold. In fact, since threshold cointegration was first introduced by Balke and Fomby (1997), it has become a feasible method for combining both nonlinearity and cointegration, and allowing, in particular, for nonlinear adjustment to long-term equilibrium, which is very suitable for studying the dynamics of market integration.

This section aims to justify and introduce the threshold cointegration approach developed by Hansen and Seo (2002). This approach defines a linear cointegration relationship as in Engle and Granger (1987), but allows the adjustment dynamics to be nonlinear and characterized by a threshold autoregressive (TAR) model, thus implying a nonlinear adjustment toward linear long-term equilibrium. This modeling framework is more robust to structural breaks, asymmetries, switching-regime and discontinuities than the usual linear cointegration (Lo and Zivot 2001).

While Balke and Fomby (1997) only consider a univariate threshold cointegration relationship with an already known cointegrating vector, Hansen and Seo (2002) make a dual contribution to the nonlinear model literature by proposing an algorithm to estimate the model parameters and a Lagrange Multiplier (LM) test for threshold cointegration. This is particularly convenient since the LM test can be computed by an ordinary least square regression involving the conventional Maximum Likelihood Estimate (MLE) of the cointegrating vector. Since the threshold is not identified under the null hypothesis, their test takes the Sup-LM form.

First, they estimate the model by the Maximum Likelihood method and execute a grid search over the two-dimensional space $(\alpha, \tau)$ in order to obtain the values of the threshold parameter and $\alpha$. They then implement a LM test for the presence of a threshold in this model. This test checks the null hypothesis ($H_0$) of "no threshold effect" against its alternative that indicates the presence of a threshold effect ($H_1$). Under $H_0$, the model is reduced to a linear Vector Error Correction Model (VECM), while under $H_1$ the model is nonlinear.

Formally, let $X_t$ be a p-dimensional I(1) time-series (*i.e.* stock price indices), which is cointegrated with one cointegrating $p \times 1$ vector $\alpha$ and let $z_t = \alpha' X_t$ be the stationary error-correction term. A linear VECM of $q+1$ order is written as follows:

$$\Delta X_t = A' X_{t-1}(\alpha) + \varepsilon_t \qquad (1)$$

Where $X'_{t-1}(\alpha) = \left(1, z_{t-1}(\alpha), \Delta X_{t-1}, \Delta X_{t-2}, ..., \Delta X_{t-q}\right)$, A is the coefficient matrix, $\varepsilon_t$ is a vector martingale difference sequence with a finite covariance matrix $\Sigma = E(\varepsilon_t \varepsilon_t')$.

In the Hansen and Seo (2002) model, all coefficients (except the cointegrating vector $\alpha$) are allowed to vary with the regimes, which may help to reproduce all regimes characterizing the integration process of stock markets. The transition between these regimes is assumed to be abrupt rather than smooth. The generalization of the threshold cointegration model developed by Balke and Fomby (1997) to the multivariate case yields the following two-regime VECM:

$$\Delta X_t = \begin{cases} A_1^{'} X_{t-1}(\alpha) + \varepsilon_t & if \ z_{t-1}(\alpha) \leq \tau \\ A_2^{'} X_{t-1}(\alpha) + \varepsilon_t & if \ z_{t-1}(\alpha) > \tau \end{cases} \quad (2)$$

where $\tau$ is the threshold parameter, $A_1$ and $A_2$ are the coefficient matrices that respectively govern the adjustment dynamics in the first and the second regime.

Finally, we should add that despite its attractiveness, this approach has never been applied to test stock market integration. Therefore, the main contribution of this article is to use the two-regime threshold cointegration approach of Hansen and Seo (2002) in order to study the short and long-term bilateral co-movements between the six major Latin American markets and the US market. The introduction of nonlinearity and threshold effects will specify an "on-off" integration process and will enable not only the extreme cases of strict segmentation and perfect integration to be reproduced but also the properties of each financial integration regime.

### 3. Data and Empirical Results

Data is monthly and consists of the S&P's IFCG total return indices for the six main emerging Latin American markets (Argentina, Brazil, Chile, Colombia, Mexico and Venezuela) and the US market, sampled over the period January 1985 to August 2005. The US market is used as a long-term target in order to check the financial integration hypothesis of the other markets toward this target. All data are obtained from DataStream International and expressed in American dollars.

Our empirical investigation involves several tests. Firstly, we apply the linear cointegration tests to check for the linkages between the markets we study. Secondly, the adjustment between these markets is checked using the threshold cointegration tests. Finally, the threshold cointegration models are estimated to reproduce the integration dynamics.

#### 3.1 Linear Cointegration Tests

The hypothesis of stationarity is required to apply linearity tests and threshold models. Thus, we firstly test for the presence of a unit root in the data using the Augmented Dickey-Fuller (ADF) test, the Philips-Perron test and the DF-GLS tests developed by Elliot *et al.* (1996). Our findings show that not all stock price series are stationary in level but are stationary in the first difference. Secondly, we check for the linear cointegration hypothesis by testing the stationarity of the residual of the cointegration relationship between the US and the other stock prices, following Engle & Granger's (1987) approach. The linear cointegration is accepted only for Brazil. This suggests that only the Brazilian stock market is integrated with the US market.[1] However, this result has to be considered carefully because the linear modeling may induce several problems of misspecification and misleading conclusions regarding cointegration when the DGP is indeed nonlinear (Lo and Zivot 2001, and Taylor 2001). To check this, we apply the more robust threshold tests, which should be more robust to nonlinearity (such as threshold effects) than linear cointegration tests.

#### 3.2 Threshold Tests

We applied threshold tests to the series for which the linear cointegration hypothesis is not accepted and we present the results in table 1. The value of the LM test of nonlinearity obtained for each country is displayed, along with the P-value obtained by a parametric bootstrap method with 5000 simulation replications. The estimation of the threshold parameters obtained by the grid search (with 300 grid points, as in Hansen & Seo 2002) is also presented.

---
[1] These results are not reported to save space, but available upon request.

*Table 1: Threshold Cointegration Tests*

| Countries | LM Test statistic | p-value | Threshold estimates $\hat{\tau}$ |
|---|---|---|---|
| **Argentina** | 14.21 | 36.4% | -1244.2 |
| **Chile** | 17.54 | 12.4% | 1921.0 |
| **Columbia** | 12.87 | 59.8% | 4203.0 |
| **Mexico** | 21.83 | 1.0% | -5597.0 |
| **Venezuela** | 12.01 | 64.2% | 840.7 |

According to these results, the null hypothesis of no threshold is rejected at 1% only for Mexico (p-value = 1.0%). Apart from Mexico, the test statistic for Chile with a value of 17.54 is the closest of all the other statistics to the rejection border at 10%. In this case, the conclusion is not so straightforward. However, for the other countries, the test statistics are too low to reject the linearity hypothesis. Thus, a threshold cointegration model is not rejected for Mexico and is not rejected also to some extent for Chile.

### *3.3 Threshold VECM*

Finally, we estimate a two-regime threshold VECM by the ML method for Mexico and Chile in order to check the threshold integration hypothesis between their stock markets and the US market. For each error-correcting model, we have two equations describing respectively the adjustment dynamics of the US index and those of the Latin American index. Each equation allows for two sets of parameters depending on the regime (1 or 2). The estimation results are presented in tables 2 and 3 for Mexico and Chile respectively. The order q = 1 of the VECM has been obtained with information criteria for different values of q up to a maximum lag order given by Schwert (1989). The numbers in brackets represent the Eicker-White standard error of the estimated coefficients. However, we have to stress that Hansen and Seo do not provide any formal distribution theory for the parameter estimates and standard errors (see Hansen and Seo, p.311). Therefore, only standard errors will be reported for these parameter estimates.

*Table 2: Threshold VECM for Mexico*

| Variable | Estimations in Regime 1 | Estimations in Regime 2 |
|---|---|---|
| **Equation 1** | | |
| $z_t$ | - 0.35 *(0.17)* | - 0.01 *(0.01)* |
| **Constant** | - 2080.5 *(1042.6)* | - 18.10 *(28.59)* |
| **dUS(-1)** | - 0.025 *(0.17)* | 0.23 *(0.08)* |
| **dMex(-1)** | 1.20 *(0.62)* | 3.60 *(0.33)* |
| **Equation 2** | | |
| $z_t$ | 0.08 *(0.04)* | 0.000 *(0.01)* |
| **Constant** | 514.97 *(260.94)* | 4.80 *(2.50)* |
| **dUS(-1)** | 0.01 *(0.042)* | - 0.01 *(0.01)* |
| **dMex(-1)** | - 0.17 *(0.23)* | 0.07 *(0.10)* |

*Note: $z_t$ is the error correction term, dUS(-q) and dMex(-q) are the first differences in the American and Mexican stock prices of order q. The numbers in brackets represent the Eicker-White standard error of the estimated coefficients.*

Overall, these results are interesting for several reasons. Firstly, for Mexico, the coefficients of the error-correction term are negative in equation 1 whatever the regime (i.e. the equation explaining the first difference of the Mexican index). For Chile, the negativity of the coefficient only occurs for equation 2 in the first regime. Secondly the t-ratios for these coefficients are not very large in the second regime for either country. However, we cannot conclude whether they are statistically significant or not as it is not possible to make any statistical inference based on Hansen & Seo's algorithm.

On the other hand, a negative sign suggests that mean reversion effects seem to appear in equation 1 for Mexico only and in the second equation for Chile, even if it is not possible to check the significance in both equations. This would imply a mean reverting mechanism between the Chilean and the American stock markets in only one regime. For Mexico, the conclusions are almost the same as for Chile.

*Table 3: Threshold VECM for Chile*

| Variable | Estimations in Regime 1 | Estimations in Regime 2 |
|---|---|---|
| **Equation 1** | | |
| $z_t$ | 0.05 *(0.02)* | 0.03 *(0.02)* |
| **Constant** | 37.12 *(9.11)* | –124.37 *(89.80)* |
| **dUS(-1)** | 0.15 *(0.10)* | 0.23 *(0.09)* |
| **dCh(-1)** | 2.12 *(0.65)* | 6.89 *(0.90)* |
| **Equation 2** | | |
| $z_t$ | - 0.01 *(0.01)* | 0.001 *(0.01)* |
| **Constant** | 2.51 *(2.37)* | 4.21 *(8.65)* |
| **dUS(-1)** | - 0.03 *(0.02)* | 0.00 *(0.01)* |
| **dCh(-1)** | 0.01 *(0.11)* | - 0.18 *(0.11)* |

*Note: $z_t$ is the error correction term, dUS(-q) and dCh(-q) are the first differences in the American and Chilean stock prices of order q. The numbers in brackets represent the Eicker-White standard error of the estimated coefficients.*

To sum up, our results suggest threshold cointegration relationships between the couples (Mexico and the US) and (Chile and the US) indicating that the Mexican and Chilean stock markets are partially integrated with the American market and showing a per-regime integration between these markets. Therefore, the dynamics of stock prices in these markets appear to depend simultaneously on local and American risk factors. More interestingly, the linkages and financial integration between these stock markets are well described and apprehended using an on-off process that is activated according to regimes, notably when stock price deviations exceed a certain threshold.

## 4. Conclusion

This paper presented a new nonlinear essay for modeling financial integration between six Latin American stock markets and the US market using the model developed by Hansen and Seo (2002). Our findings show some evidence of an on-off integration mechanism for Mexico and Chile that is activated per regime only when stock price deviations exceed a certain threshold. This suggests partial time-varying financial integration of Mexico and Chile into the US market. For Brazil, the integration process seems to follow a linear pattern, while we found no long-term relationships between the other Latin markets we studied and the US market. The approach we used in this article can naturally be extended to other emerging and developed stock markets to compare their financial integration dynamics.